\documentclass{article}
\usepackage{spconf,amsmath,epsfig,amssymb,xcolor,hyperref}

\newcommand{\proba}{\text{p}}
\DeclareMathOperator*{\argmin}{argmin}

\newcommand{\speckle}{{u}}
\newcommand{\reflectivity}{{v}}
\newcommand{\intensity}{{w}}
\newcommand{\logreflectivity}{{x}}
\newcommand{\logintensity}{{y}}
\newcommand{\logspeckle}{{s}}

\newcommand{\Emanuele}[1]{\textcolor{black}{#1}}

\title{Despeckling Sentinel-1 GRD images by deep-learning\\ and application to narrow river segmentation}
\name{Nicolas Gasnier$^\dagger$, Emanuele Dalsasso$^\dagger$, Loïc Denis$^\ddagger$, Florence Tupin$^\dagger$}
\address{$^1$LTCI, Télécom Paris, Institut Polytechnique de Paris, Palaiseau, France \\ $^\ddagger$Univ Lyon, UJM-Saint-Etienne, CNRS,
Institut d Optique Graduate School,\\ Laboratoire Hubert Curien UMR
5516, F-42023, SAINT-ETIENNE, France\\ }
\begin{document}
%
\maketitle
%

\begin{abstract}
This paper presents a despeckling method for Sentinel-1 GRD images based on the recently proposed framework ``SAR2SAR'': a self-supervised training strategy. Training the deep neural network on collections of Sentinel 1 GRD images leads to a despeckling algorithm that is robust to space-variant spatial correlations of speckle. 
Despeckled images improve the detection of structures like narrow rivers. We apply a detector based on exogenous information and a linear features detector and show that rivers are better segmented when the processing chain is applied to images pre-processed by our despeckling neural network.


\end{abstract}
\begin{keywords} SAR, denoising, deep learning, water detection, Sentinel-1
\end{keywords}
\section{Introduction}
\label{sec:intro}
Synthetic Aperture Radar (SAR) images are very useful for various fields related to Earth observation, thanks to the all-time, all-weather capability and short revisit time of a SAR system. For applications such as river monitoring that do not need interferometric information from the SLC images, Sentinel 1 GRD IW HD (Ground Range Detected Interferometric Wide swath High definition Dual polarization) images provided by ESA are very popular because of their wide availability and ease of use compared to other SAR data. 

However, they are affected by strong speckle fluctuations, with an ENL (equivalent number of looks) of 4.4. Thus, their exploitation can be challenging, as they are difficult to interpret.
In order to analyze these images, the speckle is commonly treated as a source of noise, and denoising techniques are often applied as pre-processing. Removing speckle from GRD images can be sensitive, especially given the spatially variable correlation of speckle. Thus, specific techniques need to be built to this aim. Among the state-of-the-art techniques, self-supervised deep learning approaches learn the speckle model directly from noisy images and can therefore provide results of high quality, even in presence of spatially correlated speckle where other despeckling methods produce strong artifacts.

One can ask whether such a denoising step can be beneficial even for approaches that have been designed to be robust to speckle noise. 
In this paper, we present a modified river detection method that is based on a first denoising step adapted from SAR2SAR \cite{dalsasso2020sar2sar} and a detection performed on the denoised images using a method that uses exogenous information to guide the river detection \cite{chaine}.
Section 2 presents the rationale behind the proposed method, experimental results are outlined in section 3. Finally, some conclusions are drawn in section 4.

\section{Method}
\label{sec:method}
The proposed method consists of two steps: the first step filters the speckle from the GRD images using an adaptation of SAR2SAR to Sentinel 1 GRD images. The second step segments the narrow rivers on these filtered images.

\subsection{Adaptation of SAR2SAR to GRD images}
A speckled intensity image $\intensity$ can be related to the reflectivity $\reflectivity$ through a multiplicative model
   $ \intensity = \reflectivity \times \speckle$, 
   where $\speckle$ is the speckle component.
The measured $\intensity$ is thus subject to strong fluctuations, which are signal-dependent. In order to stabilize these fluctuations, it is common to apply a homomorphic transform (\textit{i.e.}, a log transform), which turns the speckle into an additive term $\logspeckle$:
\begin{equation}
    \logintensity = \logreflectivity + \logspeckle\,,
\end{equation}
following a Fisher-Tippett distribution:
\begin{equation}
    \proba(\logspeckle) = 
    \frac{L^L}{\Gamma(L)}e^{L\logspeckle}\cdot \exp(-Le^{\logspeckle})\,,
\end{equation}
where $\logintensity$ and $\logreflectivity$ respectively correspond to the log-intensity and log-reflectivity, and $L$ is the number of looks.
Inspired by the noise2noise approach \cite{lehtinen2018noise2noise}, it is demonstrated in \cite{dalsasso2020sar2sar} that, given several pairs of noisy samples $(\logintensity',\logintensity)$ drawn under the same conditional distribution, when training a neural network using:
\begin{align}\label{eq:maximum_likelihood_loss}
    \mathcal{L}\{f_\theta(\logintensity'),\logintensity\} &= -\log \proba\left(\logintensity|f_\theta(\logintensity')\right) \nonumber \\
    & = f_\theta(\logintensity')-\logintensity+\exp{\left(\logintensity-f_\theta(\logintensity')\right)}\,,
\end{align}
the training is asymptotically equivalent to a supervised training with the (unknown) log-reflectivities $\logreflectivity$:
\begin{equation}
    \argmin_\theta\;\mathbb{E}_{X,Y|X}\left[\mathcal{L}\{f_\theta(\logintensity),\logreflectivity\}\right]\,.
\end{equation}
This makes it possible to train a network using co-registered SAR images acquired at different dates.

\Emanuele{Following the same principle outlined in \cite{dalsasso2020sar2sar}, a U-Net \cite{ronneberger2015u} network has been initially trained on noise-free reference images on which speckle noise with $L=4$ is synthetically created. In this way, pairs of GRD images are simulated in a realistic way and the network is trained only using noisy images (step \textbf{A}). In order to learn specific characteristics of GRD images, in particular the spatially-varying correlations, the network is fine-tuned on co-registered stacks of GRD images 
(step \textbf{B}). To compensate for changes occurring between images $\bf{\logintensity}_1$ and $\bf{\logintensity}_2$, the target image is replaced by $\bf{\logintensity}_2 - \hat{\bf{\logreflectivity}}_2 + \hat{\bf{\logreflectivity}}_1$. The estimates of the reflectivities $\hat{\bf{\logreflectivity}}_1$ and $\hat{\bf{\logreflectivity}}_2$ are obtained thanks to network \textbf{A}. Since network \textbf{A} was trained only on decorrelated speckle, Sentinel-1 images are downsampled by a factor 2 to reduce the impact of speckle correlations. The outputs of network \textbf{A} are then up-sampled by a factor 2 to produce estimates $\hat{\bf{\logreflectivity}}_1$ and $\hat{\bf{\logreflectivity}}_2$.
After this first fine-tuning step on real data, the network learned at step \textbf{B} is used to produce more accurate estimates $\hat{\bf{\logreflectivity}}_1$ and $\hat{\bf{\logreflectivity}}_2$ in the change compensation formula (the downsampling is no more necessary since network \textbf{B}, directly trained on Sentinel-1 images, is robust to speckle correlations). A last network is obtained at the end of this refinement: network \textbf{C}.
}
\begin{figure}
    \centering
    \includegraphics{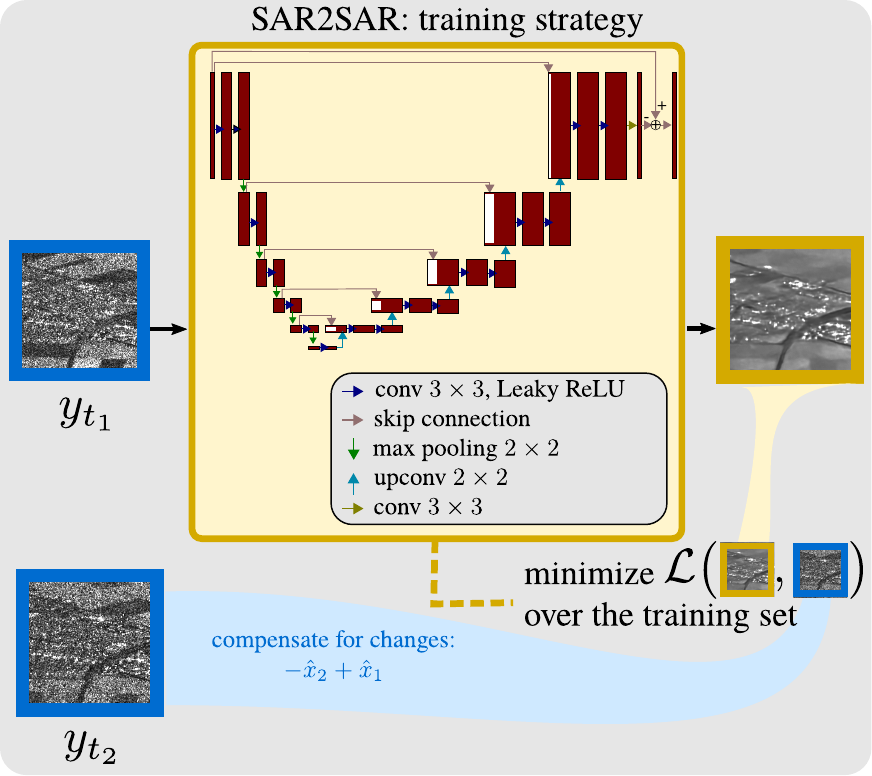}
    \caption{Summary of the self-supervised training of the SAR2SAR algorithm.}
    \label{fig:SAR2SAR_scheme}
\end{figure}

\subsection{Detection of narrow rivers in denoised GRD images}
The goal of the second step of our method is to get an accurate segmentation of the rivers in the image.
To achieve this, we adapted a framework \cite{chaine} that uses first a linear features detector and exogenous information to retrieve the river centerline. Then the river is segmented around the centerline using a specific conditional random field (CRF) approach.
The exogeneous information on the river consists of control points and can be found in prior databases such as Global River Widths from Landsat (GRWL) \cite{GRWL} in which the rivers centerlines are stored as sets of nodes.
Such information can be of a great help to distinguish rivers from other dark linear structures that are present in Sentinel-1 images, like large roads. However, due to discrepancies between the database centerline and the actual rivers on the image, both in shape and position, these river centerlines from the database cannot be directly used to segment the river, hence the need for a repositioning stage.

\subsubsection{Linear features detection on denoised GRD images}
The linear features map is computed using a linear features detector based on the Generalized Likelihood Ratio Test (GLRT) \cite{EUSAR}. 
As this detector is agnostic to the number of looks $L$, no modification was needed to adapt it
to denoised GRD images.

\subsubsection{River centerline determination as the least cost path between control points}
As the linear features detected in the previous step can correspond to actual rivers, but also to roads, relief artifacts, or even to false-positive detections caused by corner reflectors, the second stage uses exogenous information from a river database to determine the shape and the position of the river centerline in the image.

This stage is identical to its counterpart in \cite{chaine} and consists in detecting the centerline as the least-cost path between two nodes that belong to the same river in the exogenous database and are a few kilometers apart. The cost array is computed from the previously computed linear features detector response using the same parameter $N_{pow}=10$, as for noisy GRD images and with $D_{max}$ being the maximum value of $D$ in the image.

\subsubsection{River segmentation around the centerline}
The last stage of our method consists in segmenting the river around the centerline obtained in the previous step, using a conditional random field (CRF) approach adapted from \cite{chaine}, with a simplified expression.
The CRF energy is defined as the sum of a data term, a term that forces centerline pixel to be classified as water, and a regularization term that all depend on the class (land or river) of each pixel.

The data term depends on the denoised image intensity $I$ and on the class: for the river class, the data term is quadratic: 
$(\log(I)-R_{\log})^2$, where $R_{\log}$ is the estimated log-reflectivity of the water. This distribution accounts for the fluctuations of river pixel intensities caused by the remaining speckle, if any, and also for the fluctuations of the river reflectivities caused by varying roughness of the water surface.
The data term for noisy SAR images processing usually derives from a gamma distribution (to model speckle fluctuations). Variations of intensity caused by the spatial evolution of water reflectivity are considered negligible compared to that of speckle. This is no longer true when considering despeckled images. We therefore prefer a quadratic data term to account for spatial changes in water reflectivities.
In the absence of a model for the land class and in order to prevent a bias toward it, the data term for the land class is set constant and proportional to the mean value of the data term, computed on all pixels of the centerline of the river, excluding the highest values. 
We keep the same regularization term as in the method \cite{chaine}: it consists in a weighted total variation penalization where weights are inversely proportional to the magnitude of the spatial gradient of the SAR image. This penalizes transitions between river and land, except where the gradient is strong. The orientation of the gradient is also considered to align the segmentation only if the direction is right, i.e., corresponding to a river to land transition.


\section{RESULTS}
\label{sec:results}
In this section, we present the detection results we obtained by applying the proposed approach to various Sentinel 1 GRD HD images with narrow rivers.
To quantitatively assess the performance of our proposed method and to compare it with the original approach that does not use a denoising step, we compute three metrics using a manually-defined ground truth: Recall (Rec), which is the proportion of actual water pixels that are classified as water, Precision (Pre) that is the proportion of actual water among all the pixels classified as water, and F-score that is the harmonic mean of the precision and the recall.

The quantitative results are presented in table \ref{TableResults} with a comparison with the baseline method \cite{chaine} that does not use any denoising.  For each metric and each image, the best result is in bold font. The comparison shows that the proposed approach improves the detection result over the baseline for all images but one (Redon) for which the F-scores are very close.

\begin{figure*}
    \includegraphics[width=\textwidth,width=178mm]{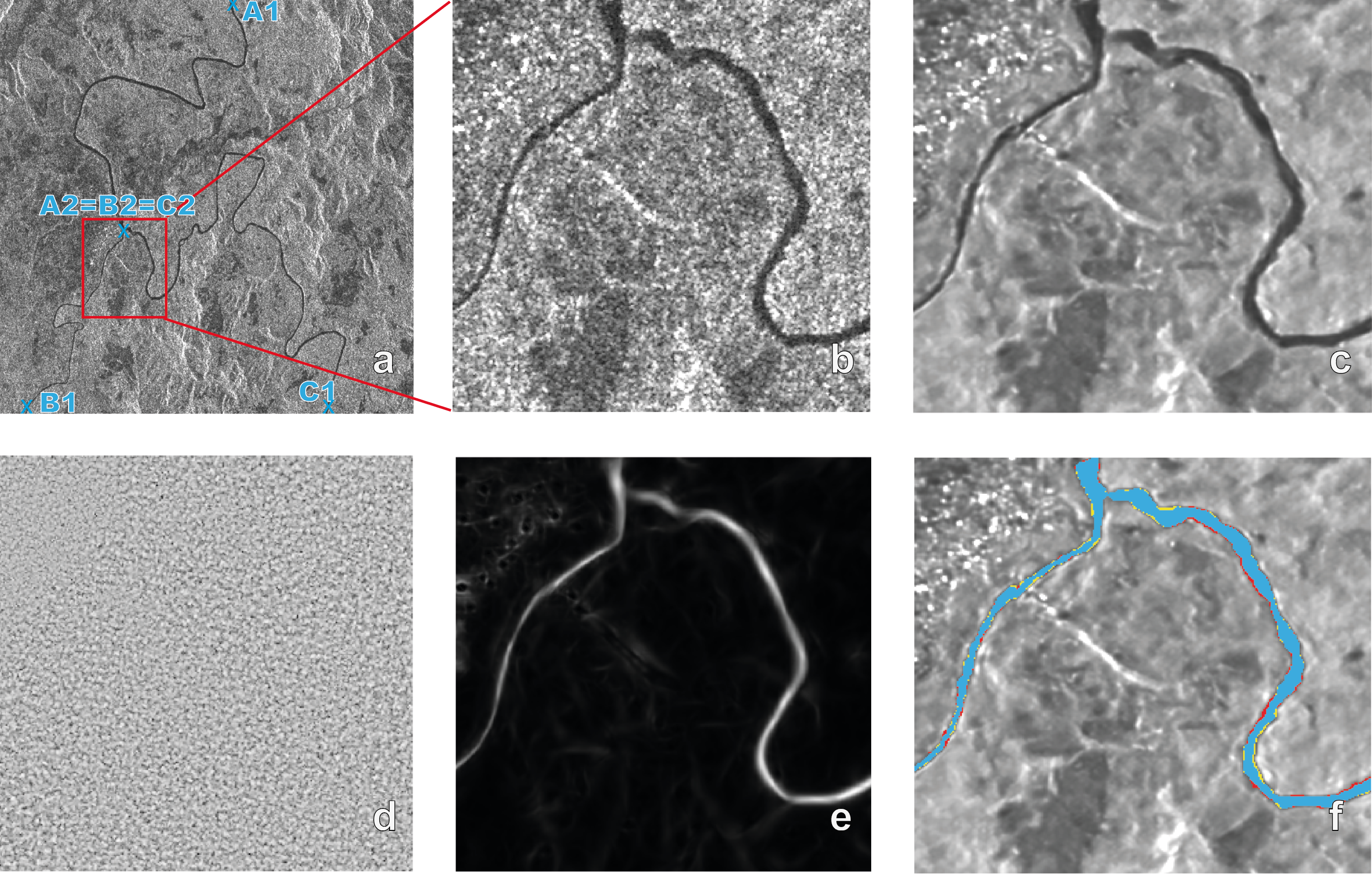}
    \caption{Some steps of the proposed method on a crop of Gaoual image. (a): Full noisy GRD image. The 3 pairs of a priori nodes (A1,A2), (B1,B2), and (C1,C2) are displayed in blue. (b-f) are cropped as presented in the red square in (a). (b): noisy GRD HD image. (c): Image denoised using the proposed denoising approach. (d): Ratio between the noisy image and the denoised image. (e) Response of the linear structures detector. (f): Result of the rivers detection. In (f), the true positive pixels are displayed in blue, the false negative in red, the false positive in yellow, and the true negative pixels as the actual denoised GRD image. The GRD images are displayed in amplitude with cropped dynamic for better visualization. The ratio image is displayed after logarithmic transformation.}
    \label{Gaoual}
\end{figure*}

The results of the different steps of the method for a crop of image 3 (Gaoual) that shows the confluence between rivers Tomine and Koumba near Gaoual (Guinea) are presented in figures \ref{Gaoual}. Figure 2.D shows the ratio between the noisy image crop and its denoised counterpart, illustrating the very good despeckling performance of the neural network: no noticeable structure is present in the ratio image and the denoised image is very smooth. 
For a more extensive analysis of the denoising results, the reader can refer to our Git repository\footnotemark.

\begin{table*}
\caption{Comparison of results between the proposed method and the baseline method}
\centering
\begin{tabular}{|c|c|c|c|c|c|c|c|c|}
\hline
   \multicolumn{3}{|c|}{Image}&  \multicolumn{3}{c|}{Baseline method} & \multicolumn{3}{c|}{Proposed Method}  \\
\hline
\#-name  & Polarization  & Size (pixels) & Pre \% & Rec\% & F-score & Pre \% & Rec \% & F-score \\
\hline
1 - Des Moines  & VH  & 1313$\times$1750 &  92.44 & 93.35 & 92.89 & \bfseries{92.54} & \bfseries{94.54}  & \bfseries{93.53}  \\
\hline
2 - Sunar & VH  & 1026 $\times$ 923 & \bfseries{82.36} & 81.71 & 82.03  & 79.12 & \bfseries{86.17} & \bfseries{82.49}\\
\hline 
3 - Gaoual & VH & 927$\times$1854 &  92.51 & 89.09 & 90.77 &  \bfseries{93.90} & \bfseries{90.00}  & \bfseries{91.91} \\
\hline
4 - Garonne & VV & 1109$\times$1704 & 97.60 & 82.44 & 89.38 &    \bfseries{97.69} & \bfseries{83.23} & \bfseries{89.89}\\
\hline
5 - Redon  & VH & 618$\times$773 & \bfseries{90.71} & 92.34 & \bfseries{91.52} &  90.41 & \bfseries{92.60}  & 91.49 \\
\hline
6 - Régina & VH & 553$\times$1216 &  89.33 & 82.95  & 86.02&  \bfseries{90.92} & \bfseries{83.98}  & \bfseries{87.31}  \\
\hline
\end{tabular}
\label{TableResults}
\end{table*}

\section{CONCLUSION}
\label{sec:typestyle}
The approach presented in this paper combines and adapts recently proposed methods to improve narrow rivers detection. First, the SAR2SAR denoising approach that was originally developed for S1 single-look images has been re-trained with S1 GRD HD time series, learning the speckle model and its non-uniform spatial correlation directly from real data. This adapted method gives denoising results of a high quality: it suppresses speckle noise while preserving fine structures without introducing notable artifacts. This is crucial for many remote sensing applications, among which narrow river detection. The images denoised with SAR2SAR-GRD are used as input for a modified narrow rivers detection method. Qualitative and quantitative experiments have shown that a preliminary denoising step significantly improves the results on the adapted river detection approach, compared to the original method working on noisy data.

In the light of these conclusions, the code of SAR2SAR-GRD is made publicly available\footnotemark[\value{footnote}], to let the remote sensing community benefit from its use to face the numerous challenges in Earth observation applications.

\footnotetext{\url{https://gitlab.telecom-paris.fr/RING/SAR2SAR}}

\bibliographystyle{IEEEtran}
\bibliography{bibliographie.bib}

\end{document}